\documentclass[article]{elsarticle}

\usepackage{hyperref}
\usepackage{graphicx}
\usepackage{mathtools}
\usepackage{multirow}
\usepackage{textcomp}
\usepackage{csquotes}
\usepackage{algorithm} 
\usepackage{algorithmic}

\DeclareGraphicsExtensions{.pdf,.png,.jpg}

\begin{document}

\begin{frontmatter}

\title{Follow Spam Detection based on Cascaded Social Information}

\author[label1]{Sihyun Jeong}
\address[label1]{Dept. of Computer Science and Engineering, Seoul National University, Gwanak-gu, Seoul 151-744, Republic of Korea}
\address[label2]{School of Electronic and Engineering, Soongsil University, Dongjak-gu, Seoul 156-743, Republic of Korea}

\ead{sihyunj@snu.ac.kr}

\author[label1]{Giseop Noh}

\ead{gsno@popeye.snu.ac.kr}

\author[label2]{Hayoung Oh}
\ead{hyoh@ssu.ac.kr}

\author[label1]{Chong-kwon Kim\corref{cor1}}
\ead{ckim@snu.ac.kr}

\cortext[cor1]{Corresponding author}

\begin{abstract}
In the last decade we have witnessed the explosive growth of online social networking services (SNSs) such as Facebook, Twitter, RenRen and LinkedIn. While SNSs provide diverse benefits – for example, forstering inter-personal relationships, community formations and news propagation, they also attracted uninvited nuiance. Spammers abuse SNSs as vehicles to spread spams rapidly and widely. Spams, unsolicited or inappropriate messages, significantly impair the credibility and reliability of services. Therefore, detecting spammers has become an urgent and critical issue in SNSs. This paper deals with Follow spam in Twitter. Instead of spreading annoying messages to the public, a spammer follows (subscribes to) legitimate users, and followed a legitimate user. Based on the assumption that the online relationships of spammers are different from those of legitimate users, we proposed classification schemes that detect follow spammers. Particularly, we focused on cascaded social relations and devised two schemes, TSP-Filtering and SS-Filtering, each of which utilizes Triad Significance Profile – (TSP) and Social status – (SS) in a two-hop subnetwork centered at each other. We also propose an emsemble technique, Cascaded-Filtering, that combine both TSP and SS properties. Our experiments on real Twitter datasets demonstrated that the proposed three approaches are very practical. The proposed schemes are scalable because instead of analyzing the whole network, they inspect user-centered two hop social networks. Our performance study showed that proposed methods yield significantly better performance than prior scheme in terms of true positives and false positives. 
\end{abstract}

\begin{keyword}
\texttt{Online Social Network, Security, Twitter, Follow Spam, Triad, Status theory}
\end{keyword}

\end{frontmatter}

\section{Introduction}

The use of social networking services (SNSs) continues to grow exponentially with the widespread adoption of smart devices such as smart phones, smart pads, smart watches, and so on. SNSs can connect people and can be used to share information in real time. SNSs such as Facebook, Twitter, and Ren-Ren are becoming the most influential mediums for building social relations, as well as for the sharing and propagation of information. According to recent announcement, Twitter, one of the largest and the most popular SNSs, passed 255m monthly active users and expects 80\% of advertising revenue from mobile users\footnote{http://thenextweb.com/twitter/2014/04/29/twitter-passes-255m-monthly-active-users-198m-mobile-users-sees-80-advertising-revenue-mobile/}

After repeated explosive growth in user population, matured SNSs such as Facebook and Twitter become a necessings in modern life in developed countries. In addition, relatively new SNSs such as RenRen and Sina Weibo, targeted for specific country or language speakers, replicate the eruptive expansion of the earlier SNSs. For example, an influential user can be exploited by a person working in online marketing to maximize the marketing effect; malicious users (attackers) disseminate false information or fraudulent messages for the purpose of phishing, scam, or malware intrusion. That is, the attackers post multiple unrelated messages with trending topics to attract legitimate users and encourage them to click the malicious links in the messages.

Spam refers to unwanted messages from unknown sources (attackers). One of the major negative aspects of SNS is spam. In the early Internet era, spam appeared in emails or SMS (short message service). However, the domain of spam expanded into SNS as the popularity and usage of the services continued to increase. False information from SNS can spread rapidly in real time. \emph{Follow spam} was reported recently and is a system that tries to increase the number of relations (or friendships) in users’ networks for the purpose of sending spam via SNS. The attack pattern of the follow spam begins with the attacker disseminating spammer accounts that follow a large number of legitimate users for the purpose of receiving a follow-back or drawing attention to the spam account \cite{ghosh2012understanding}. Due to the consequent exposure of the public to spam content, this practice definitely lowers the reliability of SNS. 

In practice, Twitter has experienced Follow spam problems, reducing users’ trust in message distribution and increasing computation overhead. In 2008, Twitter officially announced that Follow spam accounts had followed so many people that they threatened the performance of the entire system \footnote{https://blog.twitter.com/2008/making-progress-on-spam}. Even with the emerging threat from Follow spam, it has been barely investigated or researched. A contents-based spam filtering approach is employed in the Twitter spam field \cite{egele2013compa, benevenuto2010detecting, martinez2013detecting, yardi2009detecting}. However, since spam contents keep changing to avoid content-based detection by inserting URLs and images in spam messages, the contents-based approach is vulnerable against evolving message patterns. To overcome the limitations of the content-based approach, a new approach using inherent properties of SNS was introduced. 

\cite{ghosh2012understanding} first emphasized that \emph{Follow spam} should be detected by using its link-farming property. They proposed a PageRank-based ranking algorithm to lower the impact of spammers. However, this approach can be burdensome since it needs to utilize social network data for the entire network (i.e., all information for nodes and edges). Therefore, it has a high computational cost and can barely detect \emph{Follow spammers} in real time. As a result, a novel detection mechanism with low computational cost and real time spam filtering is needed while maintaining the detection performance. In this paper, we suggest two social network-based detection schemes for countering Twitter spam. First, spammer accounts are filtered out with the use of a Triad Significance Profile (TSP) that measures the structural differences between the frequencies of 13 isomorphic subgraphs. We discovered that TSP of a spammer account is different from that of a legitimate user’s account with only 2-hop social networks.  According to our experiment, 92.1\% of spammers are classified correctly when we used only TSP features for classification. This result suggests that frequency and distribution of isomorphic subgraphs could be informative features for identifying spammers. Secondly, we introduce a new detection approach using the social status (SS) theory \cite{leskovec2010signed} to distinguish spammer accounts. Legitimate users typically follow accounts of a higher status than themselves, whereas spammers are likely to follow in a random manner. With these approaches, we can confirm that cascaded social network-based approaches (TSP and SS) can effectively detect \emph{Follow spammers} with a low cost.

Our experiments on real Twitter datasets clearly show that our three mechanisms, TSP-Filtering, SS-Filtering and Cascaded-Filtering, are very practical for the following reasons. First, our approaches require only a small user-related 2-hop neighborhood social network. Actually, there are only few existing works focused on small neighborhood graph in other areas \cite{o2012identifying,akoglu2010oddball}, but none of them discovered the power of neighborhood social network clearly. Therefore, they can be applied to spam detection systems in social networks as real time solutions. Second, service providers can maintain the credibility and reliability of their SNSs by applying our approaches. Legitimate users are less likely to be blocked by the system with low false positives (5.7\%). Also, a high proportion of true positives (96.3\%) provides a secure environment to users.

Our main contributions are summarized as follows:

\begin{itemize}
\item For the first time, we discovered the feasibility of cascaded social information such as triad (or isomorphic subgraph) and social status based positive link probability as good features for classifying spammers and legitimate users in Twitter;
\item Our approaches involve more lightweight computation for real time spam detection than the previous scheme (i.e., global information). Since to check whether a certain user is spammer or not, we only focused on the 2-hop social networks of each user (i.e., local information). and extract cascaded social information from the network;
\item Based on Twitter's spam policy, novel triad graph based features and social status theory based features are proposed and cascaded together to facilitate spam detection;
\item To the best of our knowledge, our approaches are the first experiments with real world data to provide the credible and reliable Twitter system with true positive results of up to 96.3\%. We believe that our findings can provide valuable insights to the area of spam detection and defense in various social networks;
\item To sum up, this paper is the first work which clarify the difference between spammer and legitimate user in the view of subgraph consisting of neighborhood. Moreover, we suggest the novel three approaches such as TSP-Filtering, SS-Filtering and Cascaded-Filtering for the follow spam detection based on cascaded social information.
\end{itemize}

The rest of the paper is organized as follows: Firstly, we introduce interesting related works on twitter spam and spam detection mechanisms in section 2. Then, to describe our approach, we introduce the motivation for our work in section 3. Then, we propose the TSP-Filtering and SS-Filtering methods with the respective performance evaluation in section 4 and section 5. In Section 6, we propose Cascaded-Filtering with higher true positive and lower false positive than TSP-Filtering and SS-Filtering. Section 7 presents an overall evaluation and discussion of our three approaches (TSP-Filtering, SS-Filtering and Cascaded-Filtering) and Collusionrank \cite{ghosh2012understanding}.  Lastly, we close this paper with future work and the conclusion in section 8 and 9. 

\section{Related work}

\subsection{Twitter-spam Filtering}

\subsubsection{Content based Twitter-spam Filtering}

Twitter contents such as user profile, tweets, and the activity log provide various options for distinguishing spammers from legitimate users. Spammers generally write tweets that contain a hashtag and URL according to the following research studies that analyzed commonly used hashtags and URL: \cite{egele2013compa, benevenuto2010detecting, martinez2013detecting, yardi2009detecting}.

COMPA \cite{egele2013compa} detected compromised accounts that wrote spam tweets based on the tweeting language of the user's account, the tweeting time window, the URL, and the \enquote{mention} receiver. This is a personalized detection approach that learns the previous behavioral pattern of each user. Benevenuto et al. \cite{benevenuto2010detecting} and Martinez-Romo et al. \cite{martinez2013detecting} proposed classification models that learned the number of hashtags and URLs \cite{benevenuto2010detecting} or spam URLs that are used in spam groundtruth tweets. Yardi et al. \cite{yardi2009detecting} studied spammers' strategic behavioral patterns and also concluded that the use of hashtags related to trending topics is a very effective spamming strategy. Gao et al. \cite{gao2014spam} built a template based on the sentence structure of spam groundtruth tweets and used template matching to filter out spam tweets.

\subsubsection{Social-network based Twitter-spam Filtering}

As the attack strategies of Twitter spammers have become more effective, a variety of social-network-based Twitter-spam-detection approaches have been introduced.

Twitter spammers attempt to convince the public that their accounts are legitimate or famous, and they display their spam tweets to the public. It is easier for spammers to attract user interest by exploiting as much social information as possible. Previous works identified follower-market customers who purchased fake accounts that they used to follow themselves \cite{jiang2014catchsync, stringhini2013follow}. Jiang et al. \cite{jiang2014catchsync} analyzed the behavioral synchronicity among these fake accounts to detect their presence. Stringhini et al. \cite{stringhini2013follow} examined real follower markets and detected fake accounts by identifying those for which the number of followees increased suddenly but did not decrease any further. Viswanath et al. \cite{viswanath2014towards} used Principal component analysis (PCA), an anomaly-detection approach, to detect intentional \enquote{follow} or \enquote{like} behaviors from market customers.

On Twitter, \emph{Follow spam} (or link farming) refers to the act of following a mass number of people to garner attention or follow-backs\footnote{https://blog.twitter.com/2008/making-progress-spam/}. Ghosh et al. \cite{ghosh2012understanding} pointed out that most \emph{Follow spammers} attained a higher rank in existing ranking algorithms because their reciprocal-follower rate is 82\%. Based on their findings, they proposed the application of Collusionrank.

\subsubsection{Subnetwork based Spam Filtering}

The authors of \cite{wang2010don} directly crawled Twitter's data and analyzed them with both contents and social graph modeling based approaches. Based on analysis of the contents, categorized into legitimates and spams, they proved that their proposed reputation feature has the best performance among all social graph-based features for detecting abnormal behaviors. However, they only considered the relationship between outdegrees and indegrees in a simple Twitter graph for the proposed reputation feature. Even though this scheme also utilizes a small graph (subgraph), a sophisticated graph design is as only part of the triad approach. The authors of \cite{o2012identifying} used neighborhood subnetwork (i.e., ego network) to detect comment spammers in Youtube. They also utilized selected discriminating motifs and analyzed them in Youtube video-user relation network. It seems very similar with our work, but it used spam campaign related motifs. Therefore, it cannot distinguish spammers when they use other sophisticated strategy. \cite{akoglu2010oddball} extracted weighted subgraphs from target network and utilizes them as discriminating features to detect spammers. It also analyzed subgraphs by types of anomalies. Based on power law characteristic of social network,  it compared spammer to legitimate user’s neighborhood subnetwork in terms of edge or weight distribution. 

\subsection{Link spam Filtering}

\emph{Link spam} has been widely studied in the web spam detection field. This type of spam is presented as numerous links from a large number of web pages to a few target web pages. Studies on \emph{Link spam} have been receiving attention due to the limitations of PageRank \cite{page1999pagerank} and HITS \cite{kleinberg1999authoritative}. Thanks to significant link characteristics, many web link graph structure-based spam detection approaches have been introduced \cite{gyongyi2004combating, wu2005identifying, krishnan2006web, becchetti2006link, wu2006topical, castillo2007know}. TrustRank \cite{gyongyi2004combating} is one of the most popular \emph{Link spam} detection algorithms. It propagates the 'non-spam' label through social networks. Likewise, BadRank \cite{wu2005identifying} propagates the 'spam' label through social networks. Compared to PageRank \cite{page1999pagerank}, these two algorithms utilize 'non-spam' and 'spam' label propagation to lower the rank of spam webpages. \cite{benczur2006link} proposed an advanced \emph{Link spam} detection algorithm using both  'spam' and 'non-spam' label propagation. These label propagation algorithms require seed knowledge such as a set of spam nodes and a set of non-spam nodes. Therefore, noise in the initial dataset can be a critical issue for these algorithms.

\subsection{Sybil Detection}

Most SNS spam detection systems rely on Sybil detection algorithms. Peer-to-peer systems consist of multiple nodes with several connections (edges). The system has to ensure that each node is clearly identified; otherwise, a malicious user (Sybil) can attempt to create multiple fake identities masquerading as honest nodes \cite{douceur2002reclaiming}. They can then manipulate the system (by zombie machines) or attack the system in order to gain illegal profit such as positive feedback in the reputation system, getting more votes in internet polls, or targeting sites to increase their rank in Google PageRank. 

 There are two main approaches to the Sybil attack: centralized and decentralized. Centralized defense obtains admission control through a central authority. Decentralized defense has no trusted central authority and controls the IP address by binding an identity. For the decentralized attack, SybilGuard \cite{yu2006sybilguard} proposes that when each node receives $\sqrt{k}$ independent samples from a set of honest nodes of size k, a random walk can be performed to try to discover the Sybil identities by using the intersection probability between honest and Sybil groups. The number of available attack edges in Sybil are theoretically bound $O(\sqrt{k} \log ⁡k)$. SybilLimit \cite{yu2008sybillimit} is an enhanced method introduced by \cite{yu2006sybilguard}. They reduced the attack edge bound in near optimal $O(\log ⁡k)$ by exploiting various random walk methods. GateKeeper \cite{tran2011optimal} adapts the ticket distribution algorithm to obtain each node's probability of Sybil/honest users.

Secondly, the centralized method. SybilInfer \cite{danezis2009sybilinfer} assumed that the central authority knows the entire social network. After random walks, each node is assigned a Sybil/honest probability by measuring the Bayesian inference. SybilDefender \cite{wei2012sybildefender} assumed that when starting a random walk in Sybil nodes, it will pass the intersection between honest and Sybil nodes. These approaches apply community detection algorithms to find Sybil communities. SumUp \cite{tran2011optimal} addresses the vote aggregation problem by considering each voter's trust graph and calculating a set of max-flow paths from all voters.

Currently, there are many Sybil detecting methods with various social network properties. SybilRank \cite{cao2012aiding} investigates each node by assuming that honest nodes will have higher degree-normalized landing probability. A random walk is performed to measure the ranking to determine whether the account is Sybil or not. SybilShield \cite{shi2013sybilshield} utilizes a multi-community social network structure environment, considering sociological properties to cut the edge between honest and Sybil groups, performing modified random walks and figuring out the properties of multi-hop edges. SybilBelief \cite{gong2014sybilbelief} detects Sybil nodes based on a semi-supervised learning framework. This method modifies the Loopy Belief propagation system and the pairwise Markov random field to define each node's classification (Sybil/honest).

\subsection{Data Mining Approach for spam detection}

In spam detection problem, most of existing studies related the problem to classification task as follows. In general, spam classifier firstly learn features extracted from SNS using pattern of legitimate users such as the number of followees/followers, post uploading time and contents information of user profile and posts. Then, classifier determines if newly given test user is spammer or legitimate user by comparing to learned pattern. Therefore, if the test user’s behavioral pattern feature is far from legitimate user’s pattern feature (learned feature), the classifier could classify and detect the user as spammer. In some cases, classifiers adopt classification threshold to handle the tradeoff between true positive and false positive. Since reliability and credibility is crucial in using SNS, low false positive is treated particularly according to spam detection system.

In detail, \cite{kayes2015social} used linear regression for classifying and detecting spammers and it stated that deviant users from legitimate users’ patterns could be classified as spammers. Similarly, \cite{viswanath2014towards} utilized PCA (Principal Component Analysis) and it detected Facebook spammers who are distant from the principal component of legitimate users. Also, Markov random field based spam classification approach was proposed in \cite{fakhraei2015collective}. Especially, contents-based spam detection approaches largely used Naïve bayes classifier or SVM classifier with contents-related features. In the early stage of spam detection, \cite{graham2002plan,hovold2005naive} and many similar studies analyzed token or word in spam contents and applied extracted features to the Naïve bayes classifier. \cite{sculley2007relaxed} proposed optimized version of SVM spam classifier and achieved efficiency than previous ones. \cite{zhou2014cost} relieved false positive problem by adopting boundary region to classification result. Since most of spam classification is binary classification of spam and non-spam, ternary classification gives three classification labels including boundary region which means reconsidering region.

\section{Motivation}
\subsection{Web, Social Network and Twitter}

Like the Web, where the importance of each page is largely determined by who references whom, the influence of individuals on many SNSs is determined by the number of indexes they receive. For example, the number of followers is the most important factor on Twitter and determines social capital, while the number of \enquote{likes} on Facebook is similar. This feature, however, has attracted a plethora of frauds who try to increase the importance or reputation of entities by generating bogus indexes, leading to the definition of the spamdexing class of attacks. Twitter's size has expanded exponentially over the past several years and it now has over 255 million active users after a succession of rapid growth spurts that resulted in an average annual growth rate of 25\% \footnote{http://thenextweb.com/twitter/2014/04/29/twitter-passes-255m-monthly-active-users-198m-mobile-users-sees-80-advertising-revenue-mobile/}. Notably, the social-interaction structure of Twitter is very interesting. Users can follow famous persons\textendash usually celebrities or standout opinion leaders\textendash that they are unacquainted with, as well as close friends. Therefore, Twitter plays an information-propagation role in addition to the role of an online social network \cite{kwak2010twitter}.

More importantly, contrary to the Web, Facebook, and many other social networks where spam indexes usually originate from fake accounts and from a circle of colluding link farms, a malicious person can collect followers or fans from innocent, social-capital-conscious users on Twitter. Further, the high rate of follow-backs makes the detection of Twitter spammers more difficult because they receive many legitimate followers just by following target users \cite{ghosh2012understanding}. Existing link-farm-detection methods well fitted for web spam detection field therefore lose much of their effectiveness in the detection of spamdexing on Twitter.

In this paper, we demonstrate the feasibility of a cascaded SNS-based security scheme to detect \emph{Follow spam}. Different from the unpractical and heuristic approaches of previous works, with the characteristics of follow-backs we apply triad frequencies and status theory for the first time in our \emph{Follow spam} detection scheme. Note that the main purpose of this study is not the attainment of engineering optimization for the performance enhancement of prior schemes, but rather, it is the examination of the feasibility of a social-network-based security scheme in a popular online social networking site, i.e. Twitter.

Before we formalize the problem, we address the characteristics of Twitter. All 13 types of directed social graph models and social status with local information can be observed in Twitter. Additionally, Twitter has well-defined social relations in the form of the \enquote{follower} and \enquote{friend} relationships. In addition to these characteristics, spams show up frequently in Twitter. We practically exploit the policy of Twitter against spams to design our proposed scheme. Twitter's spam policy is summarized as follows:

\begin{itemize}
\item \enquote{If you have a small number of followers compared to the amount of people you are following}, the account may be considered a spam account.
\item \enquote{Multiple duplicate updates on one account} is a factor used to detect spam.
\item \enquote{If your updates consist mainly of links, and not personal updates}, it is considered spam.
\end{itemize}

First policy is related with the social-interaction structure of Twitter while second and third policies have to do with spam contents. Most previous works focused on contents analysis or full information usage of social networks with a high amount of computational overhead. Different from previous approaches considering second and third policies, we accurately detect \emph{Follow spam} using only local information of the social-interaction structure of Twitter. That is, our cascaded social network scheme is applicable regardless of the content such as Tweet, time and links.

\subsection{Link spam and Follow spam}

The concept of link farming originated from Web spam. The intent of link and \emph{Follow spam} is to increase the population of a specific (target) website or reputation. Since normal search engines (e.g., Google) place popular websites on the first page, link-farming websites create numerous links to the target website.

PageRank \cite{page1999pagerank}, the most popular website ranking algorithm, ranks websites based on the indegree of the site. Actually, the popularity of inlink nodes is also important,  but numerous inlinks are likely to increase the target website's ranking. Therefore, link farms generally contain plural links, and the links are created from many nodes to a few target nodes.

\begin{figure}[htbp]
\begin{center}
    \includegraphics[scale=0.3]{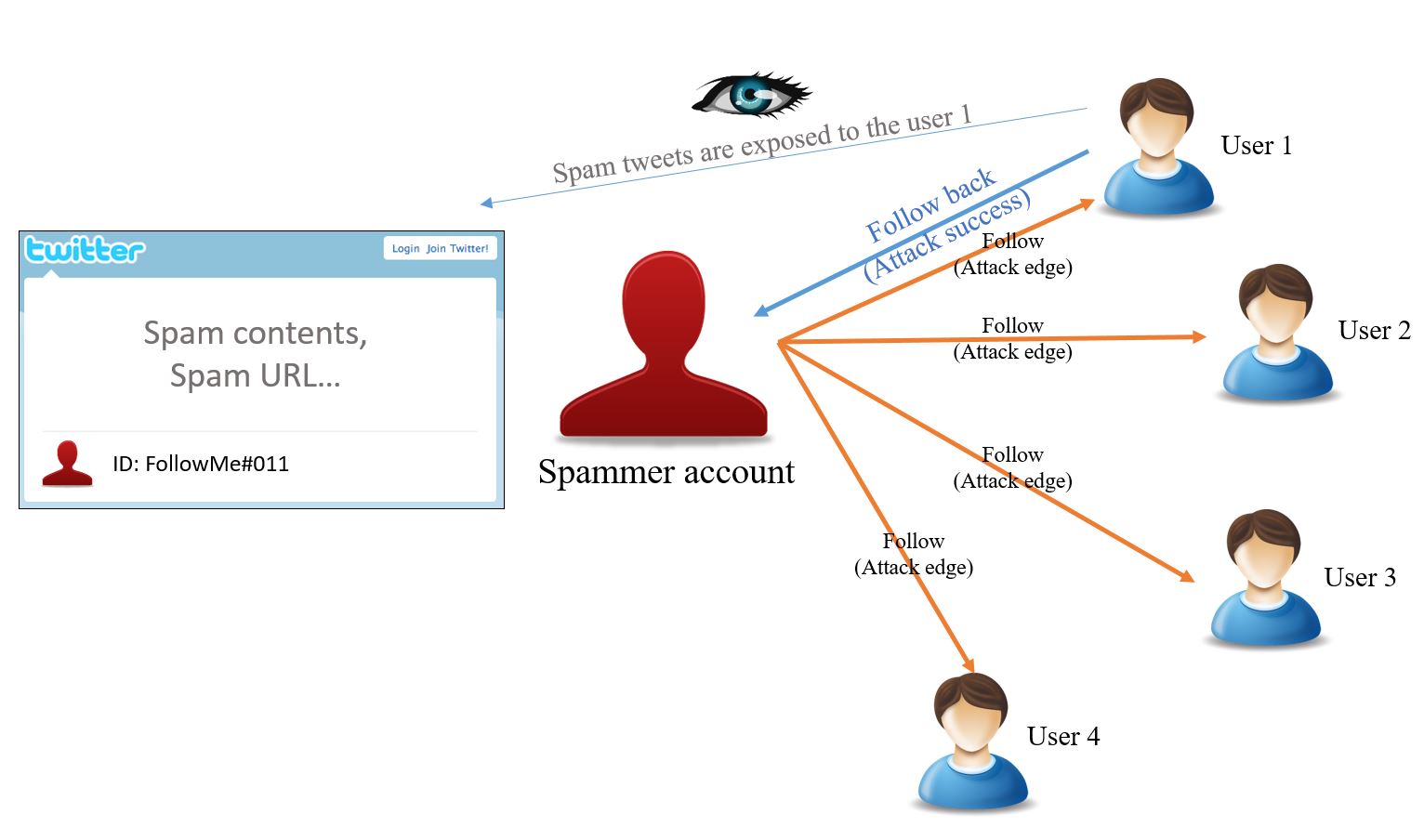}
    \caption{Overview of \emph{Follow spam}} \label{fig:fig1}
\end{center}
\end{figure}

\emph{Follow spam}, a special attack strategy on Twitter, has been shown to be a link farming technique. Figure \ref{fig:fig1} shows an example of \emph{Follow spam}.

\emph{Follow spam} consists of numerous links, but some differences exist. First, links are created by a few spammer nodes and they target many legitimate user nodes. More specifically, original link spam denotes many spammer nodes-few legitimate nodes relationship while \emph{Follow spam} denotes few spammer nodes-many legitimate nodes. Second, the purpose of \emph{Follow spam} is not just linking, but receiving a follow-back (reciprocal link). A user on Twitter can see tweets (contents) from another user when he/she follows (subscribe) the other's account. Consequently, spammers need to be followed by other users to show their spamming contents such as URL, image and advertisement.

Therefore, to gain more followers and attention,  spammers send a large number of follow links to legitimate users. Surprisingly, the majority of followers who \emph{Follow spam} accounts have been previously targeted by spam accounts. To be specific, 82\% of legitimate users send a follow-back to spammers \cite{ghosh2012understanding}. If $s$ is a spammer account and his/her outlinks are all attack edges for follow back, the attack strength of $s$ $(AS(s))$ is defined in \eqref{eq:eq1}. We defined the ratio between successful follow spam links (follow back links) of spammer $s$ (${N}_{fb}$$(s)$) and total follow spam links of $s$ (${N}_{f}$$(s)$) as $AS(s)$ as follows :

\begin{equation} \label{eq:eq1}
AS(s)=N_{fb}(s)/N_{f}(s)
\end{equation}

In \eqref{eq:eq1}, ${N}_{f}$$(s)$ has the same meaning of  “outdegree of $s$”. Therefore, the attack strength $(AS(s))$ of follow spam relies on a successful number of follow backs. 

\subsection{Twitter Dataset and Collusionrank}

We conducted an experiment with a large-scale Twitter-follow link dataset that was provided by MPI-SWS \cite{cha2010measuring}. This dataset was collected in September 2009, contains 1,963,263,821 directed social links, and the number of corresponding users is 54,981,152. We also used the \emph{Follow spammer} dataset from \cite{ghosh2012understanding} that contains 41,352 spammers as the ground truth.

Table \ref{tab:tab1} shows Twitter dataset used in our experiment.

\begin{table}[]
\centering
\caption{Twitter dataset}
\label{tab:tab1}
\begin{tabular}{cc}
\hline
\multicolumn{1}{|c|}{\textit{The number of total users}} & \multicolumn{1}{c|}{\textit{The number of spammers}} \\ \hline
\multicolumn{1}{|c|}{54,981,152}                         & \multicolumn{1}{c|}{41,352} \\ \hline
\end{tabular}
\end{table}

We compared the performance of the proposed method with that of Collusionrank \cite{ghosh2012understanding} presented in WWW 2012. Collusionrank lowers the influence scores of users who connect to spammers and filter out those users who gain high rankings by link farming. It is a user-ranking algorithm based on PageRank. Since we used the same dataset as Collusionrank, we compare the performance of the proposed method with the true positive and false positive results of Collusionrank. According to \cite{ghosh2012understanding}, Collusionrank detected 94\% of the 41,352 spammers that appeared in the last low ranking scores 10\% of ranking positions; consequently, we could extract the false positives of legitimate users (9.9\%) from Collusionrank with a detection threshold of 10\%. We reiterate that the detailed performance of Collusionrank is not described in \cite{ghosh2012understanding}, except for the true positives for spammers within the threshold of the last 10\%. Table \ref{tab:tab2} is the estimated performance of CollusionRank from a true positive value of 94\%.

\begin{table}[]
\centering
\caption{Performance estimation of Collusionrank \cite{ghosh2012understanding}}
\label{tab:tab2}
\begin{tabular}{|c|c|c|}
\hline
\textit{}       & \textit{True Positive} & \textit{False Positive} \\ \hline
Spammer         & 94.0\%                 & 6.0\%                   \\
Legitimate user & 90.1\%                 & 9.9\%                   \\ \hline
\end{tabular}
\end{table}

Collusionrank has good performance in terms of true positive and false positive, but it has some limitations as follows:

First, it needs to analyze every node and edge in a social network. The PageRank-based algorithm typically estimates every node's reputation or ranking depending on the reputation of other nodes and edge formation. However, to classify spammers, computing ranks on every node is not practical. In real SNSs, spammers disseminate spamming contents simultaneously. Therefore, a real time spam filtering approach is more effective; fast spam filtering significantly decreases the number of victims of spam.  As such, analyzing all social network information is not very pragmatic.

Second, it has a high proportion of false positives in detecting legitimate users. If 9.9\% of legitimate user accounts in Twitter were blocked, most people would stop using Twitter. A high number of true positives in detecting spammers is also crucial; but the credibility and reliability of the service are maintained by keeping the number of false positives low.

In the following sections, we propose cascaded social information-based spam detection mechanisms that overcome the limitations of Collusionrank.

\subsection{Indegree and Outdegree of Sample dataset}

Since \emph{Follow spam} has a link farming property that involves creating many outlinks, we should investigate whether spammers in Twitter have a higher outdegree than legitimate users. Also, based on Twitter's spam policy, we focus on the ratio of the indegree to the outdegree for both legitimate users and spammers.

In this paper, we use randomly selected 1,000 legitimate users and 1,000 spammers as the experimental dataset. We determined a large enough sample size with a 95\% confidence level and 5\% confidence interval.

Table \ref{tab:tab3} is the average indegree and outdegree of legitimate users and spammers.

\begin{table}[]
\centering
\caption{Average indegree and outdegree}
\label{tab:tab3}
\begin{tabular}{|c|c|c|}
\hline
\textit{}       & \textit{Average indegree} & \textit{Average outdegree} \\ \hline
Spammer         & 303.6                     & 866.5                      \\
Legitimate user & 401.5                     & 462.0                      \\ \hline
\end{tabular}
\end{table}

Inevitably, spammers tend to have approximately two times as many outdegrees as legitimate users. The most interesting observation is that the ratio between the average indegree and outdegree shows significant differences between legitimate users and spammers. The average indegree and outdegree of legitimate users are similar and the ratio between the two is 0.86. However, the ratio between the average indegree and outdegree of spammers is 0.35. This indicates that the indegree and outdegree could be roughly informative for classifying spammers.

To classify spammers by only indegree and outdegree, we used J48 and RandomForest classifiers built in Weka\footnote{http://www.cs.waikato.ac.nz/~ml/weka/}. Both algorithms are decision tree based classifiers. While J48 generates only one decision tree, RandomForest corrects overfitting problems by constructing multiple decision trees during the training process. Table \ref{tab:tab4} is the classification performance evaluation using only indegree and oudegree.

As mentioned in the Twitter spam policy, we proved that the number of outdegrees can be a highly useful feature for spam classification. However, comparison using only the number of degree types between \emph{Follow spams} and legitimate users is not enough of a performance measure to inspect spammers as shown in Table \ref{tab:tab4}. To make up for the spam detection issue, we tried to apply TSP and SS as described in section 4 and 5.

\begin{table}[]
\centering
\caption{Performance evaluation using only indegree and outdegree}
\label{tab:tab4}
\begin{tabular}{cccc}
\cline{1-4}
\multicolumn{1}{|c|}{\textit{Classifier}}           & \multicolumn{1}{c|}{\textit{Type}}  & \multicolumn{1}{c|}{\textit{True Positive}} & \multicolumn{1}{c|}{False Positive}   \\ \cline{1-4}
\multicolumn{1}{|c|}{\multirow{2}{*}{J48}}          & \multicolumn{1}{c|}{Spammer}         & \multicolumn{1}{c|}{83.9\%}                 & \multicolumn{1}{c|}{16.1\%}           \\
\multicolumn{1}{|c|}{}                              & \multicolumn{1}{c|}{Legitimate user} & \multicolumn{1}{c|}{80.7\%}                 & \multicolumn{1}{c|}{19.3\%}          \\ \cline{1-4}
\multicolumn{1}{|c|}{\multirow{2}{*}{RandomForest}} & \multicolumn{1}{c|}{Spammer}         & \multicolumn{1}{c|}{80.8\%}                 & \multicolumn{1}{c|}{19.2\%}           \\
\multicolumn{1}{|c|}{}                              & \multicolumn{1}{c|}{Legitimate user} & \multicolumn{1}{c|}{80.4\%}                 & \multicolumn{1}{c|}{19.6\%}           \\ \cline{1-4}
\end{tabular}
\end{table}

\section{Twitter-spam Detection with Triad Significance Profile}
\subsection{Follow Spam Detection with Triad Significance Profile (TSP-Filtering)}

\begin{figure}[htbp]
\begin{center}
    \includegraphics[scale=0.3]{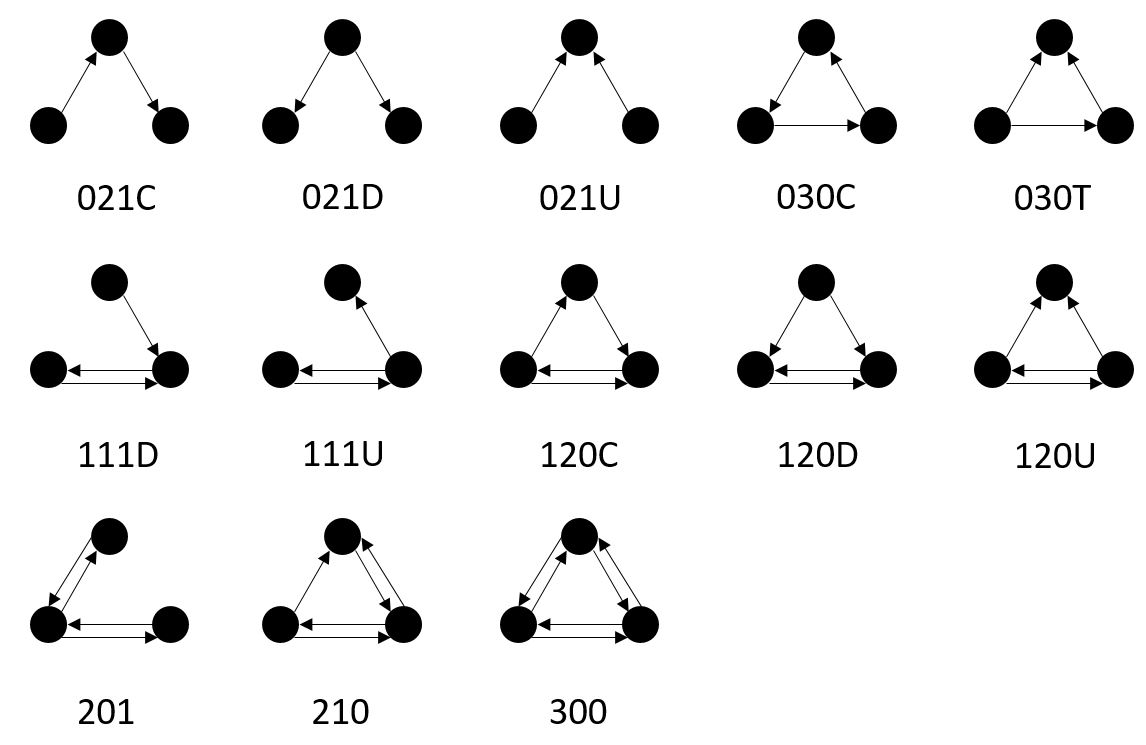}
    \caption{13 isomorphic triad classes for analyses} \label{fig:fig2}
\end{center}
\end{figure}

A prior study showed that, interestingly, several types of networks from different fields such as biology and the social sciences share common properties. In particular, \cite{milo2004superfamilies} showed that some of the 13 isomorphic triad types are over-represented while some are under-represented. To the best of our knowledge, we first used this fact to discern Twitter \emph{Follow spam}. In terms of a social graph, a user is a node and a follow from a person to another person is a directed link from the follower (the person) to the followee (another person). Figure \ref{fig:fig2} shows the 13 isomorphic triad classes introduced by \cite{wasserman1994social}.

Note that a \emph{Follow spammer} inevitably generates many follows (directed links) to receive follow-backs (redirected links). For each spammer, we found all of the corresponding triads and counted the frequency of the 13 isomorphic triad classes (for detailed representation of the triad classes, refer to Figure \ref{fig:fig2}). We performed the same procedures with legitimate users and compared the differences between the frequency of each triad class for both the spammer-centric triads and the legitimate user-centric triads. We argue that the triad frequencies of real social networks are different from those of spammers. The triad frequencies of spammers are similar to those of random networks with the same graph properties including the average indegree and the average outdegree.

\begin{figure}[htbp]
\begin{center}
    \includegraphics[scale=0.3]{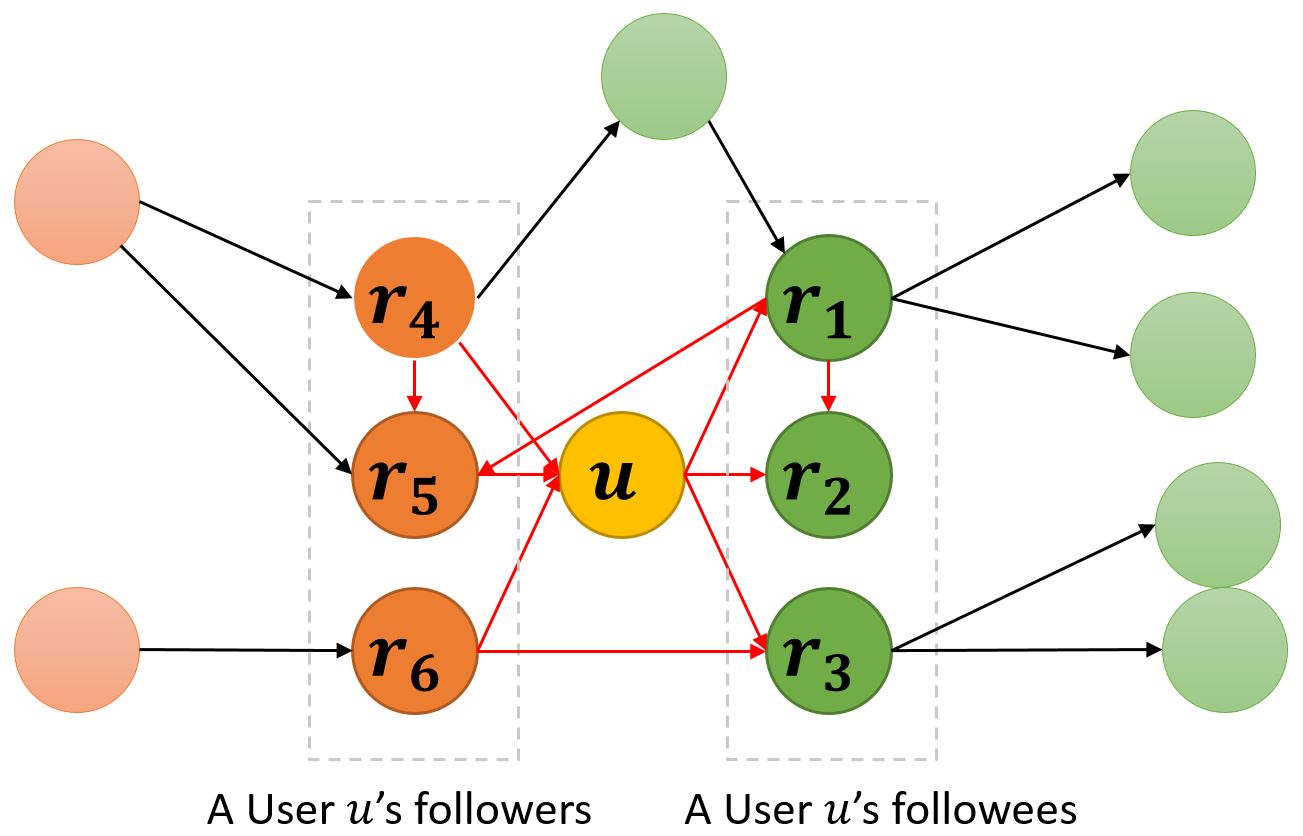}
    \caption{A user $u$'s social network graph $G_{u}$ (red-colored edges and named nodes)} \label{fig:fig3}
\end{center}
\end{figure}

For a given local network $G_{u}$ of a user $u$ as shown in Figure \ref{fig:fig3}, we estimated the number of occurrences for each triad class. $G_{u}$ consists of social links between $u$ and 1-hop neighborhoods of $u$. Suppose that $u$ is following $r_{1}$, $r_{2}$ and $r_{3}$ and is also followed by $r_{r}$, $r_{5}$ and $r_{6}$. In this case, $r_{1}$, $r_{2}$ and $r_{3}$ are \enquote{Followees} of $u$. In the same manner, $r_{4}$, $r_{5}$ and $r_{6}$ are \enquote{Followers} of $u$. Also, there are directed social links between them (represented as red-colored links in Fig. 3). To determine whether user $u$ is a spammer or not, we analyzed user $u$'s social graph $G_{u}$ consisting of 7 nodes and 10 edges. This is a subgraph of a Twitter social network, and every user can have his/her own social network.

To discover the phenomenon whereby spammer social networks comprise subgraph features that are different from legitimate user social networks, we compared spammer triad frequencies with those of legitimate users. For each triad class $i$, the statistical triad occurrence is described by the \emph{Z-score} $Z_{i}$ \cite{milo2004superfamilies} in Equation \eqref{eq:eq2}.

\begin{equation} \label{eq:eq2}
Z_{i}=(N_{spam_{i}}-<N_{legit_{i}}>)/std(N_{legit_{i}})
\end{equation}

where $N_{spam_{i}}$ is the occurrence number of the triad class $i$ in a spammer's network, and $<N_{legit_{i}}>$ and $std(N_{legit_{i}})$ are the mean and standard deviations of its appearances in the legitimate user networks, respectively. The TSP is therefore the vector of the \emph{Z-scores} that are normalized to length 1 in Equation \eqref{eq:eq3}

\begin{equation} \label{eq:eq3}
TSP_{i}=Z_{i}/(\sum {Z_{i}}^2)^\frac{1}{2}
\end{equation}

To visualize this insight from network comparison, we computed the average vector of TSP for random 1,000 spammers from the original dataset \cite{ghosh2012understanding} and normalized it. Similarly, we also computed $N_{legit_{i}}$ based on random 1,000 legitimate users. We determined that the sample size 1,000 was large enough with 95\% confidence level and 5\% confidence interval. 

\begin{figure}[htbp]
\begin{center}
    \includegraphics[scale=0.35]{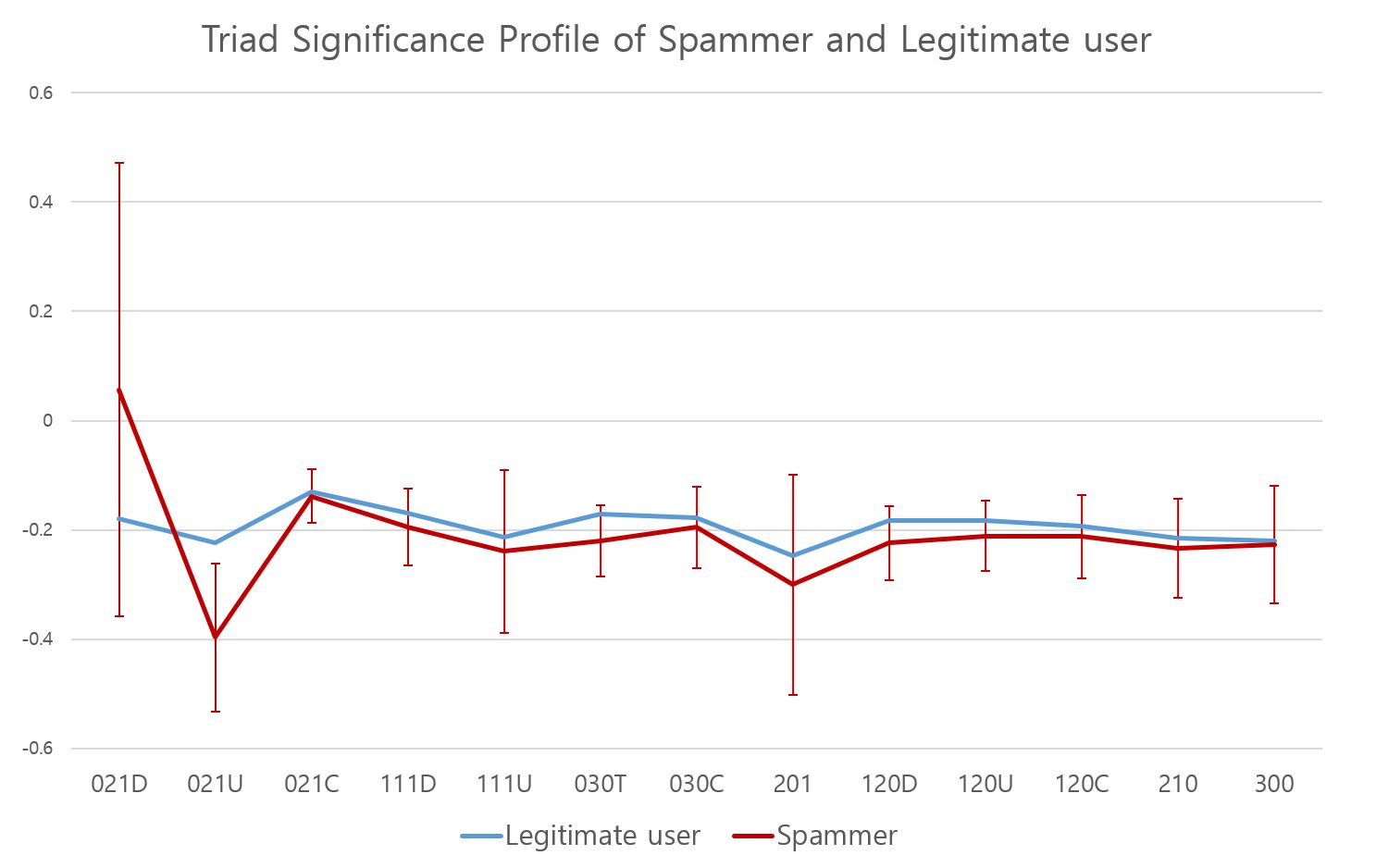}
    \caption{Average TSP of spammers and legitimate users. Error bar means standard deviation of spammers’ TSPs.} \label{fig:fig4}
\end{center}
\end{figure}

Figure \ref{fig:fig4} compares the TSPs of spammers and legitimate users. Legitimate users generally have more triads compared to spammers, meaning that the neighbors of legitimate users are socially well connected with isomorphic triad patterns; therefore, this phenomenon produced more triad counts overall. Alternatively, spammers have lower triad counts than legitimate users because their 1-hop neighbors are not likely to acquaint themselves with the other 1-hop neighbors.

Since spammers usually select their followees randomly, there are few connections between spammers' neighbors. Triad 021D, however, indicates exceptional triad counts, whereby spammers have more 021D triads than legitimate users. The 021D triad class represents the plural-following actions from a node. It also represents link-farming activity. Since the actions of \emph{Follow spammers} involve the production of numerous out-links, their high 021D triad counts make sense. The distinction between the TSPs of spammers and legitimate users therefore explains why our TSP-detection approach is feasible. We give a statistical analysis of sensitivity of every proposed strategy in discussion section (Section 7). In the following section, we provide a true-positive rate and false-negative rate to support the excellence of this method.

As mentioned earlier, we randomly sampled sets of 1,000 spammers and 1,000 legitimate users from the original dataset \cite{ghosh2012understanding} and conducted an experiment with TSP. We determined that the sample size was large enough with 95\% confidence level and 5\% confidence interval.

\subsection{TSP-Filtering}

The following process was used for the applicable value of the TSP-Filtering based on the experiment. First, we obtained the mean and standard deviations of each frequency for the triad class across all of the Twitter accounts. Since 1,000 legitimate users are sufficiently representative to support every Twitter account (confidence level: 95\%, confidence interval: 5\%), we computed the mean and standard deviations of the 1,000 randomly-sampled legitimate users. The mean value of the triad class i is $<N_{legit_{i}}>$ and standard deviation of the triad class i with $<N_{legit_{i}}>$ is $std(N_{legit_{i}})$, respectively. Figure \label{fig:fig5} shows the sampled user’s local social networks and triad frequency normalization.

\begin{figure}[htbp]
\begin{center}
    \includegraphics[scale=0.2]{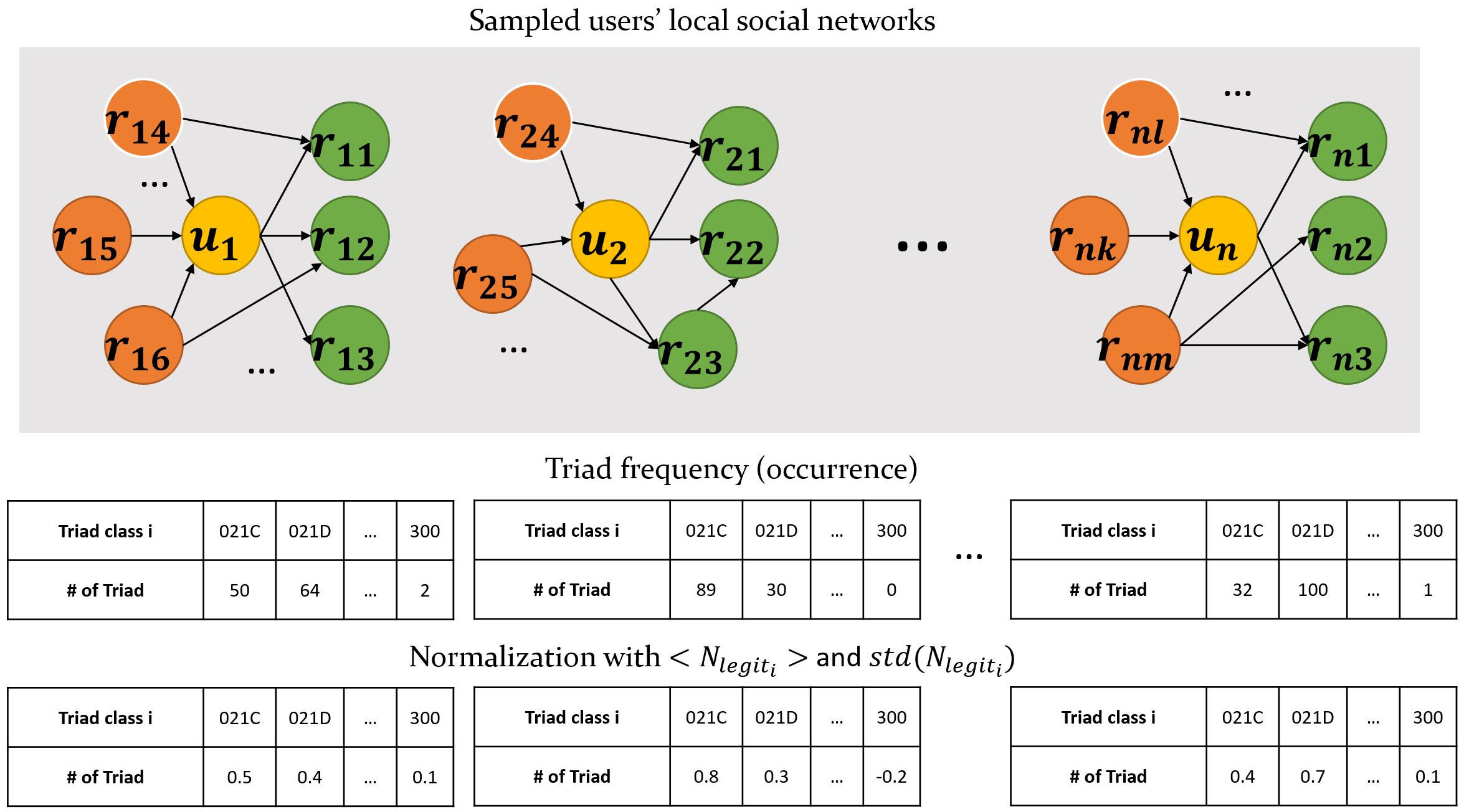}
    \caption{Triad frequency normalization} \label{fig:fig5}
\end{center}
\end{figure}

Second, we counted the spammer-triad frequencies and the legitimate user-triad frequencies for every social-network subgraph of every user account; the triad frequency represents the triad appearances in each user network. Then, we normalized the frequencies with $<N_{legit_{i}}>$ and $std(N_{legit_{i}})$ (Figure \ref{fig:fig5}). In the case of the spammer, we can use Equation \eqref{eq:eq2}; however, in the legitimate user's case, we can use the re-translated Equation \eqref{eq:eq4}, where $N_{spam_{i}}$ is the occurrence number of the triad class i in a legitimate user's network:

\begin{equation} \label{eq:eq4}
Z_{i}=(N_{legit_{i}}-<N_{legit_{i}}>)/std(N_{legit_{i}})
\end{equation}

Lastly, we computed each user's TSP using Equation \eqref{eq:eq3}. A user's TSP comprises 13 Z-scores of 13 triad classes. These 13 Z-scores could be informative for a machine-learning mechanism. We also added two features for machine learning, namely the indegrees and outdegrees of each user based on the motivation experiments.

\subsection{Performance evaluation of TSP-Filtering}

We conducted the experiment using J48 and RandomForest implemented in Weka (10-fold validation). Table \ref{tab:tab5} shows the performance evaluation results for the TSP method without indegrees and outdegrees. Table \ref{tab:tab6} shows the performance evaluation results for the TSP method with indegrees and outdegrees. On the other hand, Table \ref{tab:tab6} shows the performance evaluation results for the TSP method with indegrees and outdegrees.

\begin{table}[]
\centering
\caption{Performance evaluation using TSP-Filtering (w/o indegree and outdegree)}
\label{tab:tab5}
\begin{tabular}{cccc}
\cline{1-4}
\multicolumn{1}{|c|}{\textit{Classifier}}           & \multicolumn{1}{c|}{\textit{Type}}  & \multicolumn{1}{c|}{\textit{True Positive}} & \multicolumn{1}{c|}{False Positive}   \\ \cline{1-4}
\multicolumn{1}{|c|}{\multirow{2}{*}{J48}}          & \multicolumn{1}{c|}{Spammer}         & \multicolumn{1}{c|}{91.0\%}                 & \multicolumn{1}{c|}{9.0\%}            \\
\multicolumn{1}{|c|}{}                              & \multicolumn{1}{c|}{Legitimate user} & \multicolumn{1}{c|}{90.6\%}                 & \multicolumn{1}{c|}{9.4\%}            \\ \cline{1-4}
\multicolumn{1}{|c|}{\multirow{2}{*}{RandomForest}} & \multicolumn{1}{c|}{Spammer}         & \multicolumn{1}{c|}{92.1\%}                 & \multicolumn{1}{c|}{7.9\%}            \\
\multicolumn{1}{|c|}{}                              & \multicolumn{1}{c|}{Legitimate user} & \multicolumn{1}{c|}{91.6\%}                 & \multicolumn{1}{c|}{8.4\%}           \\ \cline{1-4}
\end{tabular}
\end{table}

\begin{table}[]
\centering
\caption{Performance evaluation using TSP-Filtering (w/ indegree and outdegree)}
\label{tab:tab6}
\begin{tabular}{cccc}
\cline{1-4}
\multicolumn{1}{|c|}{\textit{Classifier}}           & \multicolumn{1}{c|}{\textit{Type}}   & \multicolumn{1}{c|}{\textit{True Positive}} & \multicolumn{1}{c|}{False Positive}   \\ \cline{1-4}
\multicolumn{1}{|c|}{\multirow{2}{*}{J48}}          & \multicolumn{1}{c|}{Spammer}         & \multicolumn{1}{c|}{91.7\%}                 & \multicolumn{1}{c|}{8.3\%}            \\
\multicolumn{1}{|c|}{}                              & \multicolumn{1}{c|}{Legitimate user} & \multicolumn{1}{c|}{90.8\%}                 & \multicolumn{1}{c|}{9.2\%}            \\ \cline{1-4}
\multicolumn{1}{|c|}{\multirow{2}{*}{RandomForest}} & \multicolumn{1}{c|}{Spammer}         & \multicolumn{1}{c|}{92.3\%}                 & \multicolumn{1}{c|}{7.7\%}            \\
\multicolumn{1}{|c|}{}                              & \multicolumn{1}{c|}{Legitimate user} & \multicolumn{1}{c|}{92.4\%}                 & \multicolumn{1}{c|}{7.6\%}                 \\ \cline{1-4}
\end{tabular}
\end{table}

From Table \ref{tab:tab5}, even without indegrees and outdegrees, TSP-Filtering for RandomForest has a powerful spam-classification performance with 92.1\%. From Table \ref{tab:tab6}, the proposed approach with indegrees and outdegrees has 92.3\% true positives and a lower proportion of false positives (7.6\%) than Collusionrank (9.9\%). Unlike Collusionrank, which needs to analyze every link to rank every node, our TSP approach is a fast and low-cost detection mechanism that uses only the 1-hop-neighborhood network for each user. Therefore, the TSP approach is a more lightweight and efficient mechanism for detecting \emph{follow spammers} in real time.

To define a preferred sequence of attributes, we measured the importance of feature attributes based on information gain as shown in Table \ref{tab:tab7}. In Table \ref{tab:tab7}, feature attributes listed in descending order of information gain. Information gain can be computed as follows:

\begin{equation} \label{eq:eq5}
InformationGain(C,A)=Entropy(C)-Entropy(C|A)
\end{equation}

In Equation \eqref{eq:eq5}, C represents the given class such as spammer and legitimate user. A is the feature attribute. For example, \emph{InformationGain(spammer,021D)} refers to the amount of entropy decrease in a spammer class when the feature attribute 021D is provided.

\begin{table}[]
\centering
\caption{The importance of feature attributes based on information gain (TSP-Filtering)}
\label{tab:tab7}
\begin{tabular}{cc}
\hline
\multicolumn{1}{|c|}{\textit{Feature attributes}} & \multicolumn{1}{c|}{\textit{Information Gain}}\\ \hline
\multicolumn{1}{|c|}{021D}                        & \multicolumn{1}{c|}{0.2867} \\ \hline
\multicolumn{1}{|c|}{021U}                        & \multicolumn{1}{c|}{0.2556}  \\ \hline
\multicolumn{1}{|c|}{021C}                        & \multicolumn{1}{c|}{0.2366}   \\ \hline
\multicolumn{1}{|c|}{111U}                        & \multicolumn{1}{c|}{0.2267}  \\ \hline
\multicolumn{1}{|c|}{201}                         & \multicolumn{1}{c|}{0.1418}   \\ \hline
\multicolumn{1}{|c|}{030T}                        & \multicolumn{1}{c|}{0.1408}  \\ \hline
\multicolumn{1}{|c|}{111D}                        & \multicolumn{1}{c|}{0.1399}  \\ \hline
\multicolumn{1}{|c|}{120D}                        & \multicolumn{1}{c|}{0.136}    \\ \hline
\multicolumn{1}{|c|}{120U}                        & \multicolumn{1}{c|}{0.1075}   \\ \hline
\multicolumn{1}{|c|}{120C}                        & \multicolumn{1}{c|}{0.0871}    \\ \hline
\multicolumn{1}{|c|}{300}                         & \multicolumn{1}{c|}{0.0859}       \\ \hline
\multicolumn{1}{|c|}{210}                         & \multicolumn{1}{c|}{0.0794}      \\ \hline
\multicolumn{1}{|c|}{030C}                        & \multicolumn{1}{c|}{0.0465}    \\ \hline
\end{tabular}
\end{table}

As we showed in the experiment results, 021D is the most significant factor in classifying \emph{follow spammers} because of its property of two out-edges. \emph{Follow spammers} tend to have many out-edges to legitimate users. This tendency is presented naturally in 021D. The following attribute, 021U, is also significant in classifying legitimate users because of its two in-edges. Legitimate users are likely to have more followers than spammers at stable points of the Twitter SNS system. Twitter is a very special SNS due to its subscription characteristics. The more informative the users' contents are, the more followers subscribe to the user. Since most spammers upload advertisements or spamming contents on their account, they have fewer followers than legitimate users. Understandably, some legitimate Twitter users try to follow many users at the initial and transition points for subscriptions or other reasons. However, legitimate users at the steady point have a larger number of indegrees (i.e., followers) than outdegrees (i.e., friends) due to effective influence or fruitful contents because of the psychology of popularity. In addition, the remaining features of TSP are gradually reflected in the distinction between the \emph{follow spammers} and legitimate users because of the social-interaction.

\section{Twitter-spam Detection with Status Theory}

\subsection{Follow Spam Detection with Social Status (SS-filtering)}

Based on signs of directed links derived from social media sites such as Epinions, Slashdot, and Wikipedia, social status theory is first applied to predict certain kinds of social relationships \cite{leskovec2010signed}. To find strong consistency in how the model fits the data across other social networks as well as the power of influence in Twitter using our TSP filtering scheme, we additionally propose SS-Filtering for Twitter network analysis. Twitter has a special characteristic whereby users follow (or subscribe to) other users and this can be translated into the social-status theory. Generally, most Twitter users follow users who are more influential than themselves. Especially on SNS, a more-influential user is similar to a user with a high social status. When a legitimate user follows others with a higher social status, is a spammer's following pattern similar to a legitimate user? We focused on the fact that spammers are likely to follow the properties of a random network.

Our social-status-based intuition is derived from the following question: \emph{\enquote{Is the following pattern of a spammer similar to that of a legitimate user when a legitimate user follows others with a higher status?}} We focused on the fact that spammers are likely to follow the properties of a random network. In fact, spammers have little knowledge of the social relations between legitimate users. They tend to select target users randomly. One may argue that a strong spammer is able to gather information regarding the social relations of legitimate users. However, such social engineering has only partial and instantaneous influence compared to the average and apparent influence of legitimate users.

\subsection{SS-Filtering}

A social-status spam-filtering system can compute the following metric based on user \emph{u}'s 2-hop social network. To apply Twitter to the status theory, we defined the status of a user \emph{u} as the ratio of the indegree ($indegree(u)$) to the outdegree ($outdegree(u)$) of the user as shown in Equation \eqref{eq:eq6}.

\begin{equation} \label{eq:eq6}
status(u)=indegree(u)/outdegree(u)
\end{equation}

\begin{figure}[htbp]
\begin{center}
    \includegraphics[scale=0.5]{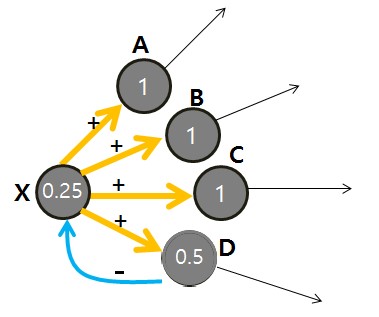}
    \caption{Social status, positive link (orange link) and negative link (blue link)} \label{fig:fig6}
\end{center}
\end{figure}

We then defined a positive link in Twitter as the probability that the user follows another user of a higher status. Figure \ref{fig:fig6} shows the concept of social status, positive link and negative link in status theory \cite{leskovec2010signed} and social balance theory \cite{cartwright1956structural, heider1944social, heider1946attitudes}. A positive link, '+' link means that a node $X$ links to each node $A$, $B$, $C$ and $D$, who has higher social status than X. On the other hand, a negative link, ‘-‘ means that the node $D$ links to a node $X$, who has lower social status than $D$.

The following Equation \eqref{eq:eq7} is the expression of the positive link probability (\emph{PLP}) of a user \emph{u}:

\begin{equation} \label{eq:eq7}
PLP(u)=N_{pos}(u)/(N_{pos}(u)+N_{neg}(u))
\end{equation}

where $N_{pos}(u)$ means the number of positive links and $N_{neg}(u)$ means the number of negative links. We assumed that negative links include links between same status. Additionally, we also consider the average status of followees. Table \ref{tab:tab8} compares social status-related features between a legitimate user and spammer.

\begin{table}[]
\centering
\caption{Average values comparison of social status-related features between legitimate user and spammer}
\label{tab:tab8}
\begin{tabular}{|c|c|c|}
\hline
\textit{Social status-related features}       & \textit{Legitimate user} & \textit{Spammer} \\ \hline
Average status of a user                      & 1.82                     & 0.39             \\ \hline
Average positive link probability (PLP)       & 0.83                     & 0.91             \\ \hline
Average status of followees ({[}0-1{]} scale) & 0.05                     & 0.0004           \\ \hline
\end{tabular}
\end{table}

\begin{figure}[htbp]
\begin{center}
    \includegraphics[scale=0.5]{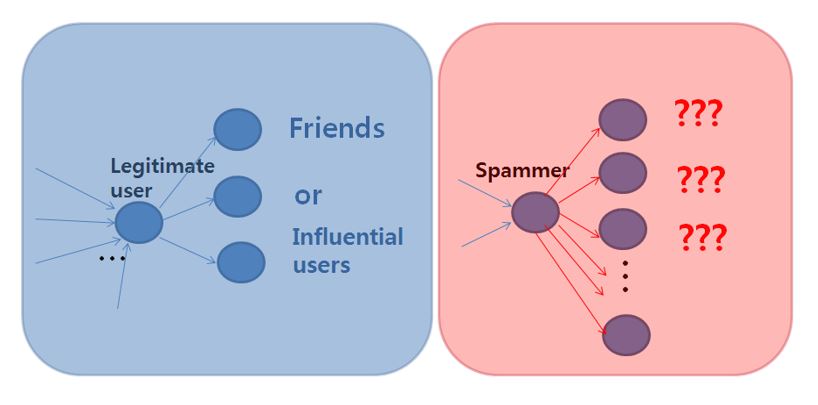}
    \caption{Comparison between legitimate users and spammers for average status of followees} \label{fig:fig7}
\end{center}
\end{figure}

We can derive interesting observations from Table \ref{tab:tab8}. First, the average status of a spammer is significantly lower than that of a legitimate user, and this is attributed to the spammer's link-farming property. In this observation of the average status of a user, the two proposed schemes (TSP filtering and SS filtering) have something in common. Second, the \emph{PLP} of a spammer to obtain a greater number of reciprocal followings is higher than that of a legitimate user. Actually, this result counters our assumption that spammers select followees randomly and the \emph{PLP} of a spammer would be lower than that of a legitimate user. However, this result is mainly due to the significantly low status of spammer accounts (i.e., a good many outdegrees of each spammer account). We arrived at this conclusion from comparison of the average status of followees. In Table \ref{tab:tab8}, we computed the average status of followees by [0-1] scale normalization for measurable comparison. Compared to the average status of legitimate users' followees, spammers' followees have definitely lower status. Therefore, though the \emph{PLP} of spammers is higher than that of legitimate users, most spammers follow users with low status. This provides sufficient evidence for our assumption. This intuitively indicates that a spammer usually targets users who are not highly influential due to lacking real social networks. Alternatively, a legitimate user follows (or subscribes to) influential users or their online friends as shown in Figure \ref{fig:fig7}.

\begin{figure}[htbp]
\begin{center}
    \includegraphics[scale=0.25]{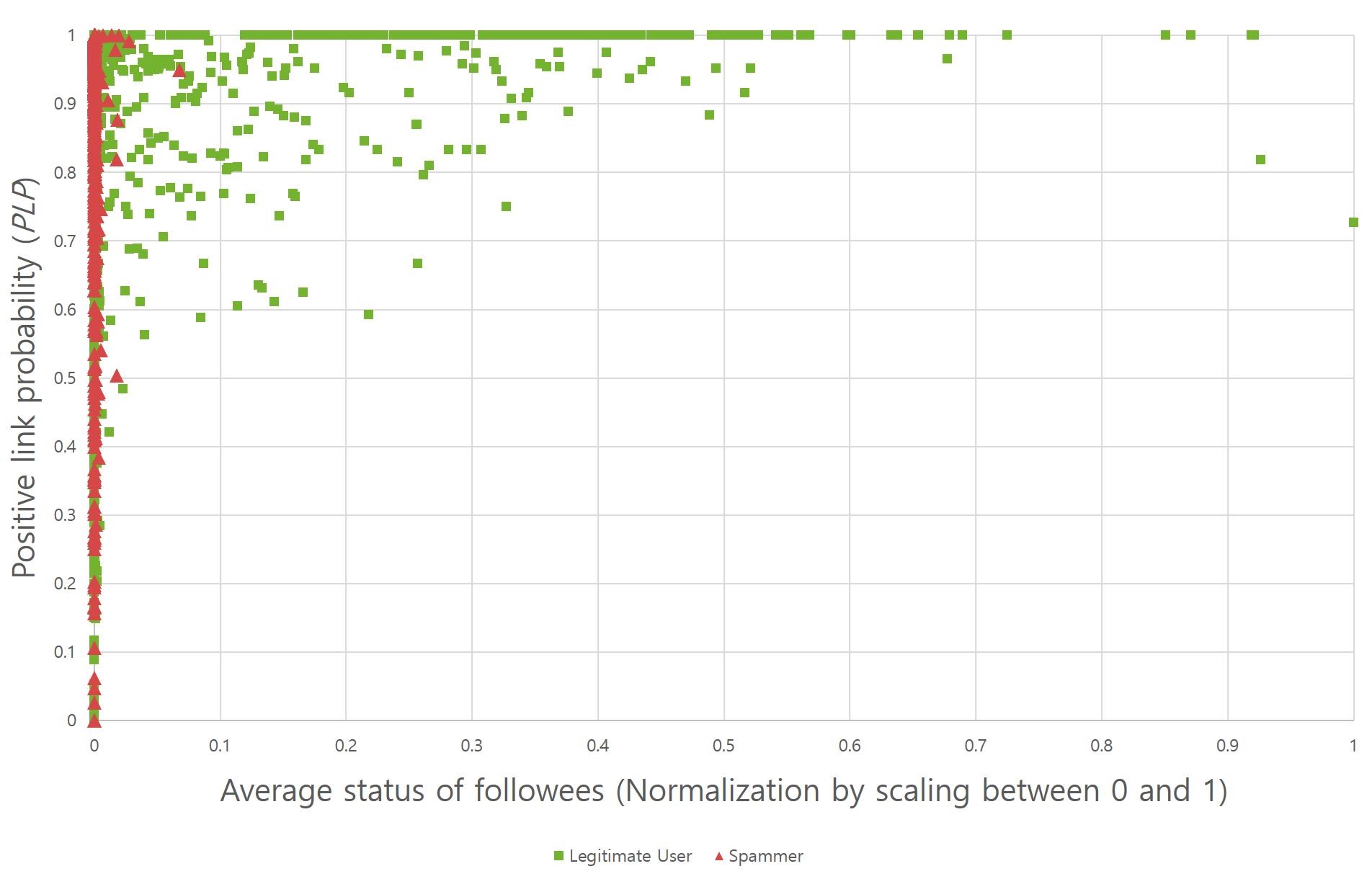}
    \caption{Green squares indicate legitimate user accounts and red triangles mean spammer accounts.} \label{fig:fig8}
\end{center}
\end{figure}

Figure \ref{fig:fig8} shows the relationship between an average followee status and the \emph{PLP}, demonstrating wide variations among the average followee statuses of legitimate users. This illustrates that legitimate users follow the properties of a real social network, whereas spammers do not.

\subsection{Performance evaluation of SS-Filtering }

For this experiment, we used the same dataset containing samples of spammers and legitimate users from the previous section. We used the user's status, average status of followees,  \emph{PLP}, indegree, and outdegree as the feature vectors. We conducted the experiment using J48 and RandomForest in Weka (10-fold validation). The following Tables (Table \ref{tab:tab9} and Table \ref{tab:tab10}) are the result of the performance evaluations of SS-Filtering considering indegree and outdegree (i.e., without or with).

\begin{table}[]
\centering
\caption{Performance evaluation using SS-Filtering (w/o indegree and outdegree)}
\label{tab:tab9}
\begin{tabular}{cccc}
\cline{1-4}
\multicolumn{1}{|c|}{\textit{Classifier}}           & \multicolumn{1}{c|}{\textit{Type}}   & \multicolumn{1}{c|}{\textit{True Positive}} & \multicolumn{1}{c|}{False Positive}   \\ \cline{1-4}
\multicolumn{1}{|c|}{\multirow{2}{*}{J48}}          & \multicolumn{1}{c|}{Spammer}         & \multicolumn{1}{c|}{90.4\%}                 & \multicolumn{1}{c|}{9.6\%}            \\
\multicolumn{1}{|c|}{}                              & \multicolumn{1}{c|}{Legitimate user} & \multicolumn{1}{c|}{82.9\%}                 & \multicolumn{1}{c|}{17.1\%}           \\ \cline{1-4}
\multicolumn{1}{|c|}{\multirow{2}{*}{RandomForest}} & \multicolumn{1}{c|}{Spammer}         & \multicolumn{1}{c|}{88.6\%}                 & \multicolumn{1}{c|}{11.4\%}           \\
\multicolumn{1}{|c|}{}                              & \multicolumn{1}{c|}{Legitimate user} & \multicolumn{1}{c|}{85.4\%}                 & \multicolumn{1}{c|}{14.6\%}                 \\ \cline{1-4}
\end{tabular}
\end{table}

\begin{table}[]
\centering
\caption{Performance evaluation using SS-Filtering (w/ indegree and outdegree)}
\label{tab:tab10}
\begin{tabular}{cccc}
\cline{1-4}
\multicolumn{1}{|c|}{\textit{Classifier}}           & \multicolumn{1}{c|}{\textit{Type}}   & \multicolumn{1}{c|}{\textit{True Positive}} & \multicolumn{1}{c|}{False Positive}   \\ \cline{1-4}
\multicolumn{1}{|c|}{\multirow{2}{*}{J48}}          & \multicolumn{1}{c|}{Spammer}         & \multicolumn{1}{c|}{88.0\%}                 & \multicolumn{1}{c|}{12.0\%}           \\
\multicolumn{1}{|c|}{}                              & \multicolumn{1}{c|}{Legitimate user} & \multicolumn{1}{c|}{90.0\%}                 & \multicolumn{1}{c|}{10.0\%}           \\ \cline{1-4}
\multicolumn{1}{|c|}{\multirow{2}{*}{RandomForest}} & \multicolumn{1}{c|}{Spammer}         & \multicolumn{1}{c|}{91.9\%}                 & \multicolumn{1}{c|}{8.1\%}            \\
\multicolumn{1}{|c|}{}                              & \multicolumn{1}{c|}{Legitimate user} & \multicolumn{1}{c|}{90.3\%}                 & \multicolumn{1}{c|}{9.7\%}                 \\ \cline{1-4}
\end{tabular}
\end{table}

Consequently, the proposed approach also shows a good proportion of true-positives (91.9\%) and a lower amount of false positives (9.7\%) compared to Collusionrank. Similar to the TSP method using the 1-hop-network of each user, the SS method uses only the small 2-hop-network for each user. These lightweight schemes make the spam-filtering process much faster. Considering that Collusionrank uses every link and node for computation, our detection mechanism using the social status is as efficient as the TSP approach in real time spam filtering. Table \ref{tab:tab11} shows the importance of feature attributes based on information gain with Equation \eqref{eq:eq5}.

\begin{table}[]
\centering
\caption{The importance of feature attributes based on information gain (SS-Filtering)}
\label{tab:tab11}
\begin{tabular}{|c|c|}
\hline
\textit{Feature attribute} & \textit{Information Gain} \\ \hline
A user's status            & 0.386                     \\ \hline
Followee's status          & 0.327                     \\ \hline
PLP                        & 0.107                     \\ \hline
\end{tabular}
\end{table}

In SS-Filtering, the user's status is the most significant feature similar with TSP-Filtering. This is because spammers have fewer indegrees than outdegrees. Therefore, the status of a spammer is lower than that of a legitimate user. However, the followee's average status was also important as well as the user's status. Normally, a legitimate user subscribes to informative users' accounts, and the status of these users is likely to be high due to their many followers. This means that legitimate users follow users with higher status than themselves. On the other hand, spammers tend to select and follow legitimate users randomly. As a result, their followees (targets) may have a lower status than themselves.

\section{Twitter-spam Detection with Cascaded approach (Cascaded-Filtering)}

In previous sections, we proposed TSP-Filtering and SS-Filtering using partial information (i.e., up to the 2-hop social network of a user) for lightweight and real-time spammer detection. Both algorithms have fewer false positives than Collusionrank, but their true positives are not superior to Collusionrank. Therefore, we suggest a hybrid approach (Cascaded-Filtering) that utilizes every feature attribute used by both TSP-Filtering and SS-Filtering. We conducted an experiment on each of 1,000 legitimate user and 1,000 spammer account with TSP features (TSP of 13 triad classes), social status features (the user's status, average status of followee, \emph{PLP}), indegree and outdegree. Table \ref{tab:tab12} shows the performance evaluation results using Cascaded-Filtering.

\begin{table}[]
\centering
\caption{Performance evaluation using Cascaded-Filtering}
\label{tab:tab12}
\begin{tabular}{cccc}
\cline{1-4}
\multicolumn{1}{|c|}{\textit{Classifier}}           & \multicolumn{1}{c|}{\textit{Type}}   & \multicolumn{1}{c|}{\textit{True Positive}} & \multicolumn{1}{c|}{False Positive}   \\ \cline{1-4}
\multicolumn{1}{|c|}{\multirow{2}{*}{J48}}          & \multicolumn{1}{c|}{Spammer}         & \multicolumn{1}{c|}{94.0\%}                 & \multicolumn{1}{c|}{6.0\%}            \\
\multicolumn{1}{|c|}{}                              & \multicolumn{1}{c|}{Legitimate user} & \multicolumn{1}{c|}{92.4\%}                 & \multicolumn{1}{c|}{7.6\%}            \\ \cline{1-4}
\multicolumn{1}{|c|}{\multirow{2}{*}{RandomForest}} & \multicolumn{1}{c|}{Spammer}         & \multicolumn{1}{c|}{96.3\%}                 & \multicolumn{1}{c|}{3.7\%}            \\
\multicolumn{1}{|c|}{}                              & \multicolumn{1}{c|}{Legitimate user} & \multicolumn{1}{c|}{94.3\%}                 & \multicolumn{1}{c|}{5.7\%}                 \\ \cline{1-4}
\end{tabular}
\end{table}

\section{Discussion}

\subsection{Overall Performance Comparison}

In this paper, we compared three \emph{Follow spam} filtering mechanisms (TSP-filtering, SS-filtering and Cascaded-Filtering) with Collusionrank. Collusionrank is the first \emph{Follow spam}-targeted filtering algorithm published. It is a PageRank-based algorithm, so it can be applied when the spam-filtering system contains information on every Twitter social network. We propose the TSP-filtering and SS-filtering methods, both of which can be applied with only the 2-hop social network of a user; this is the most powerful feature of these methods. Table \ref{tab:tab13} shows the overall comparison of the three methods with Collusionrank.

\begin{table}[]
\centering
\caption{Overall performance comparison}
\label{tab:tab13}
\begin{tabular}{|c|c|c|c|c|}
\hline
\textit{}                                                                  & \textit{Collusionrank} & \textit{\begin{tabular}[c]{@{}c@{}}TSP-\\ Filtering\end{tabular}} & \textit{\begin{tabular}[c]{@{}c@{}}SS-\\ Filtering\end{tabular}} & \textit{\begin{tabular}[c]{@{}c@{}}Cascaded-\\ Filtering\end{tabular}} \\ \hline
\begin{tabular}[c]{@{}c@{}}True Positive\\ (Spammer)\end{tabular}          & 94\%                   & 92.3\%                                                            & 91.9\%                                                           & 96.3\%                                                                 \\ \hline
\begin{tabular}[c]{@{}c@{}}False Positive\\ (Legitimate user)\end{tabular} & 9.9\%                  & 7.6\%                                                             & 9.7\%                                                            & 5.7\%                                                                  \\ \hline
\end{tabular}
\end{table}

Both TSP-filtering and SS-filtering have lower true-positive rates and more favorable false-positive rates. However, we are convinced that our detection mechanism is more effective for the following reasons. In general, for SNSs such as Twitter and Facebook, real-time spam detection is the most important issue. Since spammers simultaneously disseminate numerous spamming contents to SNSs, fast and immediate filtering with minimal information is needed to prevent spamming. To detect spam contents promptly, a lightweight computational cost is essential. TSP-filtering and SS-filtering identify spammers by using small subgraphs with up to 2-hop social networks for a user. On the other hand, Collusionrank requires more than 24GB of RAM to perform computation on a dataset with 1,963,263,821 edges. Therefore, for application to the entire Twitter-user population, it is not efficient to use every node and edge. We therefore suggest a divide-and-conquer approach like our TSP-filtering and SS-filtering methods for immense SNSs.

Another issue is the false-positive rate. For the reliability and convenience of SNSs, a spam-filtering system should not filter legitimate users as spammers, since this blocks users' utilization abilities and leads to notoriety and systemic failure. In comparison with the false-positive rate, compensation of the true-positive rate of TSP-filtering and SS-filtering is natural and easier. Spam-reporting services are incorporated into the design of most SNSs for the detection of content abusers, and this complementary tool could be helpful for the detection of subtle spamming actions. Accordingly, TSP-filtering and SS-filtering could be practical spam-filtering mechanisms for use under SNS conditions.

The performance of Cascaded-Filtering, which employs every feature attribute used in TSP-Filtering and SS-Filtering, is superior to the other three schemes including CollusionRank. This scheme can accurately detect more spammers and block or suspend less legitimate users. Consequently, this result supports the idea that there is a distinction between the legitimate user's social network and the spammer's social network. Especially, this scheme does not require the full social network information, including users that are not directly related to the account, to determine whether a specific user is a spammer or not. Moreover, for the service provider, half the number of false positives of Collusionrank is pretty attractive to ensure a reliable and convenient SNS system.

\begin{figure}[htbp]
\begin{center}
    \includegraphics[scale=0.35]{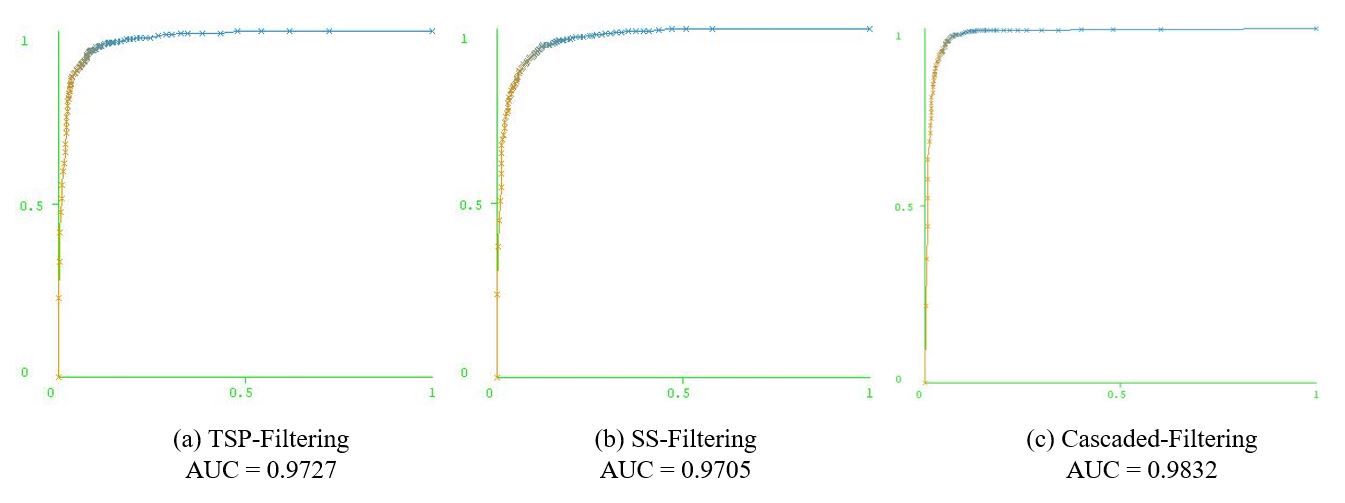}
    \caption{ROC curve of (a) TSP-Filtering and (b) SS-Filtering (c) Cascaded-Filtering
(X axis : False positive , Y axis : True positive)
} \label{fig:fig9}
\end{center}
\end{figure}

Figure \ref{fig:fig9} shows the Receiver Operating Characteristic (ROC) curve and the Area Under the ROC Curve (AUC) for every proposed approach. Due to its high true positive and low false positive values, Cascaded-Filtering has the highest AUC. To compare with Collusionrank, we estimated the AUC of Collusionrank using its true positive and false positive values. Consequently, compared to Collusionrank (AUC=0.92), our approaches are competitive and require much less social network information.

\subsection{Sensitivity Analysis}
\begin{figure}[htbp]
\begin{center}
    \includegraphics[scale=0.35]{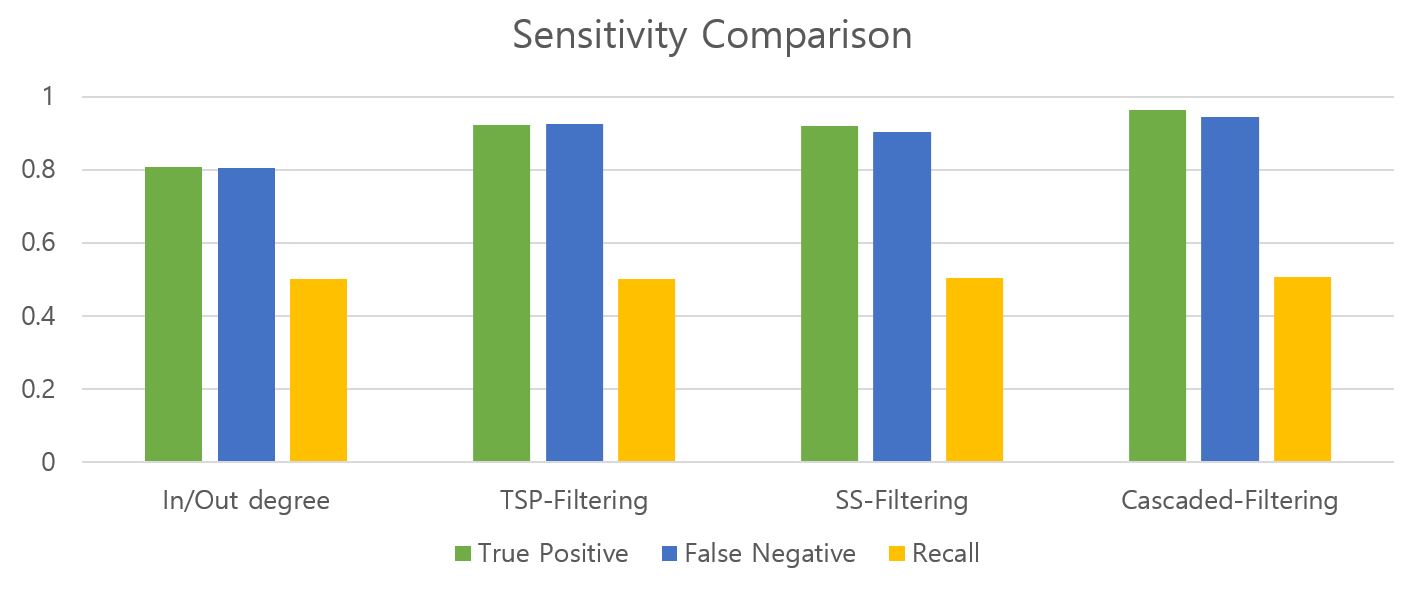}
    \caption{Sensitivity of the proposed detection strategies} \label{fig:fig10}
\end{center}
\end{figure}

Figure \ref{fig:fig10} shows a statistical analysis to evaluate the sensitivity of the TSP-detection strategy. 

\begin{table}[]
\centering
\caption{True Positive, False Negative and Recall of every suggested strategy}
\label{tab:tab14}
\begin{tabular}{|l|l|l|l|}
\hline
                   & True Positive & False Negative & Recall \\ \hline
In/Out degree      & 0.808         & 0.804          & 0.501  \\ \hline
TSP-Filtering      & 0.923         & 0.924          & 0.500  \\ \hline
SS-Filtering       & 0.919         & 0.903          & 0.504  \\ \hline
Cascaded-Filtering & 0.963         & 0.943          & 0.505  \\ \hline
\end{tabular}
\end{table}
We computed the recall(sensitivity) as following equation \label{eq:eq8}:

\begin{equation} \label{eq:eq8}
recall=TruePositives/(TruePositives+FalseNegatives)
\end{equation}

To sum up, from table \ref{tab:tab14}, two proposed approaches, SS-Filtering and Cascaded-Filtering, show better sensitivity than spam classification using In/Out degree only. Actually, TSP-Filtering has slightly lower sensitivity than those approaches, it counters this limitation with both high true positive and false negative. Cascaded-Filtering, which is the hybrid approach using every feasible features, shows the best performance at sensitivity with superior true positive and false positive. 

\subsection{Complexity Analysis}

Following pseudo-code is TSP-Filtering and it consists of two steps: Triad census algorithm and Triad Significance Profile computation algorithm. At first, we modified input of Triad census algorithm \cite{batagelj2001subquadratic} to apply to two-hop input subnetwork. $G=(V,E)$ is the two-hop directed subnetwork of a user. 

\begin{algorithm}
\footnotesize
	\caption{TriadCensus}
	\label{al1:triadcensus}
\begin{algorithmic}
\STATE $INPUT$ : $G$
\FOR{$i$:=1 to 16}
\STATE $N_{i}=0$
\ENDFOR
\FOR{$v$ $\in$ $V$}
	\IF{$u$ $\in$ $v$}
		\STATE{$S$:= $Neighbor(u)$ $\cup$ $Neighbor(v)$ $\setminus$ \{$u$,$v$\} } 
		\IF{Link($u$,$v$) and Link($v$,$u$)}
			\STATE TriType:=3
		\ELSE
			\STATE TriType:=2
		\ENDIF
		\STATE  $N$[TriType]:=$N$[TriType] + $n$ - $|S|$ - 2
		\FOR{each $w$ $\in$ $S$}
			\IF{$u < w$ $\vee$ ($v < w \wedge w < u \wedge \neg Link($v$,$w$)$ }
				\STATE TriType:=TriType[Tricode($v$,$u$,$w$)]
				\STATE $N$[TriType] := $N$[TriType]+1
			\ENDIF
		\ENDFOR
	\ENDIF 
\ENDFOR
\STATE $sum$:=0
\FOR{$i$:=2 to 16}
	\STATE $sum$:=$sum$+$N$[$i$]
\ENDFOR

\STATE $N$[1]:= (1/6)$n$($n$-1)-$sum$
\STATE return $N$

\end{algorithmic}
\end{algorithm}

Algorithm 1 computes frequencies of 16 isomorphic triad types. In our work, we used only 13 isomorphic triads because triad types with isolated nodes are counted frequently so that this interrupts observing significance of triads. From this algorithm, we can get frequencies of 13 isomorphic triad types of given graph G. TSP-Filtering get 2-hop neighborhood social network of a certain user as input. This algorithm has the complexity O($m$) and $m$ is the number of edges in the graph G. In algorithm 1, TriType means how many nodes are connected each other. This indicates that an isolate node exists when TriType is 2. \cite{batagelj2001subquadratic} used the concept of Tricode to count triad efficiently, but we don’t explain detail of the concept in this paper.

\begin{algorithm}
\footnotesize
	\caption{Triad Significance Profile}
	\label{al1:tsp}
\begin{algorithmic}
\STATE $INPUT$ : $N$, $<N_{legit_{i}}>$, std($N_{legit}$)

\FOR{$i$:1 to 13}
	\STATE $Z_{i}$=($N[i]$-$<N_{legit_{i}}>$)/std($N_{legit_{i}}$)
\ENDFOR
\FOR{$i$:1 to 13}
\STATE $TSP_{i}$=$Z_{i}$/($\sum$ ${Z_{i}}^2)^\frac{1}{2}$)
\ENDFOR
\STATE return $TSP$

\end{algorithmic}
\end{algorithm}

Algorithm 2 computes Z-score and TSP of a certain input user. For input attributes, this algorithm has input vectors consisting of mean and standard deviation of legitimate users’ triad frequency. These attributes could be got at preprocessing phase, which computes mean and standard deviation of legitimate users for the training value. Since our work is to classify spammers from legitimate users, we should compare the user’s triad frequency $N$ (testing value) to legitimate user’s triad frequency $N_{legit}$ (training value). This algorithm has time complexity O(1) because of a simple normalization process. Therefore, TSP-Filtering is a combined algorithm of TriadCensus algorithm and TSP algorithm. The novelty of our work is that we used a user’s 2-hop neighborhood social network as initial input. Also, we compare each user’s census to overall legitimate users to observe overrepresentation and underrepresentation in each isomorphic triad. Overrepresentation and underrepresentation of triads could be the distinct features to classify spammers from legitimate users. So, TSP-Filtering has O($m$) time complexity for a target user.
Following pseudo-code is SS-Filtering. Actually, in preprocessing phase, we can compute the status of a user with his/her indegree and outdegree. Since we defined the status of a user as the ratio of the indegree to the outdegree of the user (Equation \eqref{eq:eq6}), we just update three values – indegree, outdegree and status – in real time case. So, we only need positive link probability ($PLP$) of a user. The simple algorithm for $PLP$ is as follows. $Status$ means a table with status of every user in $G$.

\begin{algorithm}
\footnotesize
	\caption{Positive Link Probability}
	\label{al1:plp}
\begin{algorithmic}
\STATE $INPUT$ : $G$, $Status$

\STATE $PositiveLink$:=0
\FOR{each $u$ $\in$ $Neighbor(v)$}
	\IF{$Status[v]<Status[u]$}
		\STATE $PositiveLink$:=$PositiveLink$+1
	\ENDIF
\ENDFOR
$PLP$:=$PositiveLink$/$|Neighbor(v)|$
\STATE return $PLP$

\end{algorithmic}
\end{algorithm}

Algorithm 3 computes the ratio of the number of neighbors with higher status than user v to the number of neighbors of user $v$. Since SS-Filtering highly depends on preprocessing phase, this simple algorithm has the complexity O($n$) when $n$ means the number of users in $G$.

\begin{table}[]
\centering
\caption{Complexity analysis}
\label{tab:tbl15}
\begin{tabular}{|c|c|c|c|c|}
\hline
                 & Collusionrank & TSP-Filtering & SS-Filtering & \begin{tabular}[c]{@{}c@{}}Cascaded-\\ Filtering\end{tabular} \\ \hline
Time complexity  & O(N+M)        & O(m)          & O(n)         & O(m+n)                                                        \\ \hline
Space complexity & O(N)          & O(m)          & O(n)         & O(m+n)                                                        \\ \hline
\end{tabular}
\end{table}

Additionally, Table  \ref{tab:tab15} shows the complexity analysis of the proposed schemes and Collusionrank. Actually, since our work is for each user’s 2-hop neighborhood social network, complexity comparison with Collusionrank seems to be ironic. To sum up, Collusionrank is for classifying large number of spammer accounts from whole social network, but TSP-Filtering, SS-Filtering and Cascaded-Filtering is for determining if a user is spammer or not. n means the number of user (nodes) in 2-hop neighborhood social network and m is the number of relation (edges) in 2-hop neighborhood social network. Also, N means the number of every users (nodes) in whole social network and M means the number of every relations (edges) in whole social network. This table showed that our approaches need less computation than Collusionrank.

\section{Future Work}

\subsection{Dynamic follow spammer detection}

First of all, camouflage attack, which uses compromised user account to disseminate spam contents, could be one of the interesting future work.  To solve the camouflage attack problem, unsupervised scheme can be considered for detecting newly occurred spammers to the proposed scheme (i.e., unsupervised scheme based on cascaded social information). For example, we can select self-organizing map (SOM) among machine learning algorithms for mapping between spammer labeled with supervised scheme based on cascaded social information and new spammer based on unsupervised scheme. 

In detail, SOM is one of clustering algorithms using Euclidean distance concept between source and destination. If we train spammer patterns with cascaded social information scheme as a supervised concept that can be a partial part of pre-defined map of SOM for processing and clustering new input automatically as a unsupervised concept. SOM calculates Euclidean distance between spammer account with cascaded social information and new Twitter account to check whether the tendency of new account is closer to spammer or legitimate. In each epoch, SOM can update the tendency of each account based on Euclidean distance dynamically and distinguish between time-varying spammer and legitimate based on clustering map. In future, we will apply SOM or other optimal solution for detecting dynamic spammer according to time.

\subsection{Generalized applications of the proposed scheme}

In the view of social behavior, people have different social networking style based on their purpose of using SNS. For example, spammers make numerous following links to unspecified individuals with the aim of spreading spam contents. On the other hand, if someone wants to make friends in SNS, they would carefully select users as friends based on a personal preference and closeness. And finally, these social links made by various users could be interpreted in cascaded social information.

As shown in our experiment, cascaded social information and its structural analysis suggest various future application in social network security. In future work, we are to find a correlation between various behavior patterns of using SNS and cascaded social information. The behavioral patterns of using SNS can be more specified by various types regardless of spammer and legitimate. One of the types is a false rumor in SNS. Since false rumor is unverified and unconfirmed information, it seems very attractive sometimes according to its degree of sensationalism. In general, false rumors are spread widely in the community and proven to be false eventually. However, we need to focus on users who spread false rumors in the view of their behavioral patterns. If someone has high social status in SNS, he/she tends to hesitate to spread rumors because he/she is anxious for impairing his dignity or credibility by spreading misinformation. Therefore, we could guess that false rumors are spread by users who aren’t related to social credibility or reputation. In conclusion, false rumor detection based on cascaded social information could be one of great topics for future generalized application.

\section{Conclusion}

On Twitter, one of the most popular social networking services (SNSs), a new kind of spamming strategy has emerged known as Follow spam. The goal of this paper is to classify follow spammers by utilizing social network properties in the individual’s local social network. To solve this problem, we proposed three novel cascaded social information based spam detection mechanisms (TSP-Filtering and SS-Filtering) and a hybrid approach (Cascaded-Filtering). 
These approaches analyze and exploit social network properties such as Triad Significance Profile (TSP) and Social Status (SS). We conducted large-scale experiments on real Twitter datasets. The results from analyzing individual-related small local social networks support our assumption that a spammer’s social network is different from a legitimate user’s social network. We compared our approaches to Collusionrank, the PageRank-based representative algorithm of the follow spam detection. Cascaded-Filtering was found to be the most competitive and superior approach while requiring much less social network information. 
In conclusion, with a high proportion of true positives (96.3\%) and low amount of false positives (5.7\%), our approaches are very secure and practical mechanisms that can be applied as real time spam detection systems.

\section*{Acknowledgments}
This work was supported by the National Research Foundation of Korea(NRF) grant funded by the Korea government(MSIP) (No. NRF-2015R1A2A1A01007400), Institute for Information \& communications
Technology Promotion(IITP) grant funded by the Korea government(MSIP) (No. B0190-15-2017, Resilient/Fault-Tolerant Autonomic Networking Based on Physicality, Relationship and Service Semantic of IoT Devices) and Basic Science Research Program through the National Research Foundation of Korea(NRF) funded by the Ministry of Science, ICT \& Future Planning(NRF-2014R1A1A1003562).

\section*{References}

\bibliography{cascaded-spam-detection}

\end{document}